# The epistemological status of Astrobiology: a problematic case of integration of scientific disciplines

Juan Campos Quemada
Universidad Complutense de Madrid

**Abstract**

Astrobiology is a scientific discipline that studies life in the Universe. We call it a discipline and not a science because some authors have cast doubts over its epistemological status by calling it "a science without an object of study". As with astrophysics, the scientific nature of astrobiology is related to historical-narrative sciences and nomothetic sciences. This discipline also integrates complex methodological and conceptual problems which originate from the methodological and epistemological differences that exist between physics and biology. This is why it is so important to evidence the different philosophical approaches from which its results are interpreted.
After a brief historical introduction, we will consider the problem of life and we will analyse the influence that different philosophical approaches have on astrobiology. Subsequently, we will introduce ontological and epistemological questions that originate from interdisciplinarity, for example, their role in a physicalistic type of reductionism and in teleology.

(Keywords: Astrobiology, interdisciplinarity, transdisciplinarity, scientific integration, exobiology, life)

**Resumen**

La astrobiología es la disciplina científica que estudia la vida en el Universo. Decimos disciplina y no ciencia porque su estatuto epistemológico ha sido puesto en duda por algunos autores que la denominan "ciencia sin objeto de estudio". Al igual que la astrofísica, la naturaleza científica de la astrobiología está relacionada tanto con las ciencias de tipo histórico-narrativo como con las ciencias nomotéticas. Además, esta disciplina incorpora problemas complejos de tipo metodológico y conceptual derivados de las diferencias metodológicas y epistémicas entre la física y la biología. Por eso, es importante evidenciar los distintos enfoques filosóficos desde los que se interpretan sus resultados.
Tras una breve introducción histórica, plantearemos el problema de la vida y analizaremos las influencias que las distintas posiciones filosóficas ejercen sobre la astrobiología. Posteriormente trataremos de forma preliminar cuestiones ontológicas y epistemológicas derivadas de la interdisciplinaridad, como el papel que desempeñan en el reduccionismo de corte fisicalista y la teleología.

(Palabras clave: Astrobiología, interdisciplinaridad, transdisciplinaridad, integración científica, exobiología, vida)

> *"Hay infinitos mundos, sean semejantes o desemejantes; pues siendo los átomos infinitos, como poco ha demostramos, son también llevados remotísimamente. Ni los átomos (de los cuales se hizo o se pudo hacer el mundo) quedaron todos sumidos en un mundo ni en infinitos (mundos); ni semejantes a éste, ni desemejantes. Así, no hay cosa que impida la infinidad de los mundos"*[1]

### 1. La astrobiología: heredera de una tradición

La cuestión sobre la existencia de vida fuera de la Tierra no es ni mucho menos una cuestión de la postmodernidad. Presenta una genealogía que puede retrotraerse hasta la conocida posición de los pitagóricos sobre la similitud de la Luna y la Tierra: incluso en lo que respecta a los hipotéticos pobladores de la primera. Esta tradición fue recogida por Plutarco quien nos propone lo que se puede entender como una de las primeras teorías exobiológicas:

> *"...yerran quienes consideran que la Luna es un cuerpo tórrido e ígneo. Además, quienes interpretan que los seres vivos de allí puedan tener la misma disposición que los de aquí en lo tocante a la existencia, el alimento y el modo de vida, parecen ignorar las diferencias que presenta la naturaleza, en cuyo seno es posible hallar variaciones y desigualdades de mayor entidad entre los respectivos seres vivos que entre estos y los cuerpos inanimados."*[2]

Observamos como hace dos mil años se formulaban con precisión hipótesis como las condiciones de habitabilidad, diversidad, etc. que pertenecen al actual programa de investigación en astrobiología.

La hipótesis de los mundos habitados renació en los albores del Renacimiento. Las líneas dedicadas a las posibilidades de vida en otros mundos no escaparon a la imaginación, al intelecto y a la pluma de Nicolás de Cusa:

> *"no es cognoscible por el hombre si la región de la Tierra es más perfecta en grado o más innoble con respecto a las regiones de las demás estrellas, que la del Sol o la de la Luna y las restantes. Tampoco con respecto al lugar, porque este lugar del mundo sea habitación de los hombres, animales y vegetales, que son más innobles que los habitantes de la región del Sol y de otras estrellas. Pues aunque Dios sea el centro y circunferencia de todas las regiones de las estrellas, y procedan de Él todas las distintas naturalezas de las noblezas, en ninguna región deja de haber habitantes y no hay ningún lugar de los cielos ni de las estrellas que esté vacío, y no parece ser sólo esta tierra la habitada por las cosas menores. Sin embargo, por la naturaleza intelectual que habita en esta Tierra y en su región no parece que pueda darse más*

---

[1] Epicuro, *Carta a Herodoto*, apud. Diógenes Laercio. *Vidas, Opiniones y Sentencias de los filósofos más ilustres*. pp.1362-1363
[2] Plutarco. *Sobre la cara visible de la Luna*. Gredos. Madrid 2004. 940b

*noble y más perfecta según esta naturaleza, aunque haya habitantes de otro género en otras estrellas.*

*Así pues, el hombre no apetece otra naturaleza, sino sólo ser perfecto en la suya. No hay proporción, pues, entre los habitantes de otras estrellas, cualesquiera que sean, y los de este mundo"[3]*

La naturaleza de los hipotéticos seres que pueblan otros mundos nada tiene que ver con la nuestra. Esta afirmación la suscribe Lord Martin Reeves:

> *"They could be staring us in the face and we just don't recognise them. The problem is that we're looking for something very much like us, assuming that they at least have something like the same mathematics and technology,"[4]*

Aunque parece que el antiguo presidente de la Royal Society no estaría de acuerdo con Nicolás de Cusa en fundamentar su argumentación en que la omnipotencia de Dios es la hacedora de la vida y, por tanto, decide a su antojo el cuándo, el dónde, el cómo y el qué de su acto creador.

El inicio de lo que se denomina astronomía moderna se desarrolla principalmente en dos periodos: en el primero, la visión copernicana del mundo construye un nuevo modelo matemático donde encuentra acomodo la cosmología aristotélica con una nueva astronomía. La teoría heliocéntrica formulada en el *De Rebolutionis Orbium Coelestium* de Nicolás Copérnico será sólo el primer paso en la consecución de un nuevo modelo del mundo que se consolidará gracias a la precisión y pulcritud experimental del astrónomo Tycho Brahe, la extraordinaria capacidad matemática de Johanes Kepler, y las incipientes observaciones telescópicas realizadas por Galileo Galilei.

La controversia astronómica sobre los sistemas del mundo ptolemaico y copernicano quedó apagada hasta el año 1609 cuando se produce la publicación de la *Astronomia Nova* por Johanes Kepler y las primeras observaciones telescópicas de Galileo Galilei, que finalmente llegan a la imprenta con el título de *Sidereus Nuncius* en 1610.

En la obra de Kepler se postulan las dos primeras leyes del movimiento de los cuerpos celestes: los planetas se mueven en órbitas elípticas en torno al Sol, que ocupa uno de sus focos, y los planetas en su movimiento orbital barren áreas iguales en tiempos iguales, es decir, los planetas se mueven más rápido cuando su órbita está más cerca del Sol que cuando está más lejos. Esta

---

[3] Nicolas de Cusa. *De Docta Ignorantia.* Barcelona. Orbis, 1984.p.130
[4] Reeves, M. [http://www.telegraph.co.uk/science/space/7289507/Royal-astronomer-Aliens-may-be-staring-us-in-the-face.html] 20.05.2010

es la primera piedra de la modernidad para construir un modelo matemático, fisicalista y determinista del mundo.

En el *Sidereus Nuncius,* Galileo muestra la naturaleza cambiante y corrupta del mundo supralunar: informa sobre la naturaleza y morfología terrenal de la Luna, anuncia el descubrimiento de los satélites mediceos que giran alrededor de Júpiter, describe la composición estelar de la conocida como Vía Láctea y estudia la morfología extraña de Saturno y las fases de Venus.

Si tenemos en cuenta estos resultados y tomamos en consideración la demostración de la mutabilidad de los cielos tras la aparición de varios cometas y el descubrimiento de una nueva estrella (nova) en 1604, la naturaleza sublunar de los planetas y sobre todo, el argumento más sólido a favor del heliocentrismo: la existencia de cuerpos celestes que se movían en torno a un planeta y no, como defendía el corpus tradicional, alrededor de la Tierra, el cuerpo conceptual cosmológico aristotélico quedó herido de muerte. Durante el siglo XVII se encendieron disputas astronómicas, cosmológicas y teológicas como refleja el *Dialogo sopra i due massimi sistemi del mondo tolemaico e copernicano* (1632) de Galileo, debates que no se apagaron hasta que la obra de Newton, *Philosophia Naturalis Principia Matemática (1687),* construida sobre la física matemática, experimental, arquimediana o galileana, ahogó definitivamente los argumentos de la física aristotélica y medieval.

En este ambiente de cuestionamiento del mundo antiguo, Giordano Bruno fue uno de los encargados de emprender la demolición de las jerarquías cosmológicas y ontológicas que "atrapaban" la imaginación del hombre desde Aristóteles y Platón. Su vehemente defensa de la infinitud del Universo, sus mundos y sus habitantes, se interpretan como paradigma de la 'apertura de mente del científico' aunque tal vez se pueda afirmar que tiene más que ver con luchar desde el fanatismo contra el fanatismo:

> *"Nosotros no estamos en la circunferencia de ellos más de lo que ellos están en la nuestra; ellos no son para nosotros el centro más de lo que nosotros somos para ellos; nosotros pisamos nuestra estrella y somos abarcados por nuestro cielo no de otro modo que ellos por el suyo"*[5]

Respecto al acaloramiento de Bruno, contrasta la tibieza de Galileo sobre la posibilidad de habitabilidad de los mundos si tenemos en cuenta la actitud abiertamente positiva de su colega Johanes Kepler que tras conocer el descubrimiento de los satélites de Júpiter afirmó:

---

[5] Bruno, G. *Sobre el infinito universe y los mundos.* Aguilar. 1972. p. 71

*"no es inverosímil que haya habitantes no sólo en la Luna, sino también en el mismo Júpiter, ahora que por primera vez se desvelan estas regiones [...] si hay cuatro planetas girando en torno de Júpiter con diversas distancias y periodos, ¿cuál es su finalidad, pregunto, si no hay nadie en el globo de Júpiter que capte con su mirada tan admirable variedad?"* [6]

La cuestión de la vida y la habitabilidad de los mundos gozan definitivamente de la promoción más autorizada cuando Imanuel Kant escribe la *Historia general de la naturaleza y teoría del cielo*:

"La perfección del mundo espiritual como la del material, crece y progresa en los planetas desde Mercurio hasta Saturno, o quizá más allá de él (si es que existen más planetas), en justa graduación en la proporción de sus distancias al Sol […] En síntesis todo concuerda para confirmar la ley citada. Sin duda alguna, la naturaleza ha extendido sus reservas con mayor abundancia en el lado opuesto del universo" [7]

Como podemos deducir de lo expuesto hasta el momento, la astrobiología representa una apuesta de la ciencia actual dentro de una larga tradición de conjeturas e imaginación en torno a las viejas preguntas por la existencia de vida en otros mundos y nuestra posición en el cosmos.

---

[6] Kepler, J. *El mensajero sideral*. Alinaza Editorial. 2007. p.127
[7] Kant, I. *Historia general de la naturaleza y teoría del cielo*, p. 199

## 2. La astrobiología como disciplina científica

La astrobiología es la disciplina científica que estudia la vida en el Universo[8]. Decimos disciplina y no ciencia porque algunos autores[9] dudan de su cientificidad argumentando que una de sus subdisciplinas como la exobiología – la biología no terrestre - es una "ciencia sin objeto de estudio" pues hasta la fecha de redacción de este trabajo no se han encontrado evidencias de vida extraterrestre.

Podemos comprender la supuesta paradoja de la exobiología bajo dos puntos de vista: el del astrofísico que estudia constructos teóricos hipotéticos durante décadas sin haberlos observado – agujeros negros, materia oscura no bariónica, las supercuerdas, etc. y bajo la larga tradición filosófico-mítico-religiosa que hemos narrado y que presenta antecedentes históricos tan ilustres como la panspermia[10].

La Unión Astronómica Internacional utiliza el nombre de bioastronomía que goza de poco predicamento fuera de esta institución, situación que comparte con el término cosmobiología, aunque, tal vez, éste sería el sustantivo más adecuado para la nueva ciencia.

Si para aclarar nuestras ideas buscamos astrobiología en el diccionario de la Real Academia, nuestro esfuerzo será infructuoso: no figura como término admitido en el DRAE. Entonces se nos puede ocurrir bucear en el diccionario para encontrar la definición de un término con raíz griega "ἄστρον" que posea cierto aire de familia con la astrobiología:

> *"Astrofísica: Parte de la astronomía que estudia las propiedades físicas de los cuerpos celestes, tales como luminosidad, tamaño, masa, temperatura y composición, así como su origen y evolución"*[11]

Parece evidente que si cambiamos la palabra física por biología y sus conceptos y propiedades, la definición no concuerda con la utilizada en este trabajo.

> *"Parte de la astronomía que estudia las propiedades biológicas de los cuerpos celestes, tales como la reproducción, metabolismo, especiación, mutación, así como su origen y evolución"*

---

La referencia anterior pretende mostrar que, a pesar de la analogía con la física en cuanto a la construcción del término, la astrobiología es un neologismo que no encuentra en el "βιος" el mismo fundamento metafísico y epistémico que la astrofísica halla en la "φύσις".

Por otro lado, si preguntamos a los astrobiólogos sobre su ciencia, nos describen la astrobiología como la disciplina que conecta la astrofísica – ciencia que estudia el contenido físico (materia y energía) del universo, incluido su origen (cosmología) – y la astroquímica – ciencia que estudia el contenido molecular del universo y su evolución en distintos entornos - con la biología para la comprensión del contenido biológico del Universo.

Menos preciso se muestra el promotor y primer director del Centro de Astrobiología (CAB) Juan Pérez Mercader:

> *"La astrobiología es una nueva ciencia que surge de la necesidad de investigar el origen, presencia e influencia de la vida en el Universo. Es una rama del conocimiento relativamente reciente, pues su punto de partida se puede situar en 1998, cuando la NASA creó el NASA Astrobiology Institute.*
>
> *La astrobiología es, desde su mismo origen, transdisciplinar. Relaciona ciencias tales como la astronomía, la astrofísica, la biología, la química, la geología, la informática, la antropología y la filosofía, entre otras. La esencia del estudio astrobiológico es el análisis de problemas científicos desde el punto de vista de varias disciplinas independientes con sus propios métodos y aproximaciones.*
>
> *Esto es especialmente útil en el caso de fenómenos históricos como la vida, en los que subyacen bases simples como la física y la química que se manifiestan de forma compleja como la biología.*
>
> *No hay una definición consensuada de astrobiología, aunque su campo de interés es perfectamente reconocible: además de todo lo que tiene que ver con la comprensión del fenómeno de la vida tal y como lo conocemos (su emergencia, condiciones de desarrollo, adaptabilidad -extremofilia-, etc.), también involucra la búsqueda de vida fuera de la Tierra (exobiología) y sus derivaciones, como son la exploración espacial o la planetología."[12]*

Desde el título del exordio, *La vida como consecuencia de la evolución del universo,* el profesor Pérez Mercader se sitúa claramente en lo que respecta a la teleología del programa de investigación antes de poderlo definir, aunque "su campo es reconocible" en la línea de aquél

---
[12] Pérez Mercader, Juan. *La vida como consecuencia de la evolución del universo*. Centro de Astrobiología (CAB)[http://cab.inta.es/astrobiologia.php?lng=es.] Madrid. 15.03.2010

historiador que era capaz de reconocer el arte cuando lo veía a pesar de su incapaz para definirlo[13]. Por el contrario, no queda nada clara su posición respecto a la reducción y a la emergencia de la complejidad.

El astrobiólogo Bruce Jakosky se pregunta sobre el estatuto científico de la Astrobiología como ciencia e indaga en su naturaleza histórica. Afronta el problema desde dos posiciones filosóficas: el *falsacionismo popperiano* y la sociología de las revoluciones científicas y los paradigmas *kuhnianos*.

El director del equipo de la Universidad de Colorado concluye:

*"The relevance of philosophy of science to the practice of science emerges in a number of the issues that have been brought out here: the difficult of disproving or falsifying hypotheses in the real world, the lack of the objective means by which to evaluate and compare competing hypothesis, the resulting idea of a cultural influence on scientific results, the ability for honest and sincere scientists to disagree about the interpretation of measurements or observations, the idea of the acceptance of scientific hypotheses by consensus, and the role of fundamentally new in allowing the development of brand new hypothesis and paradigms. […] Especially important is the role that exploration science plays in astrobiology and an understanding of how this approach to doing science differs from the more traditional views of science"*[14]

*Es evidente la dificultad epistémica que aborda el programa astrobiológico cuando uno de sus promotores tiene que pedir honestidad a los científicos a la hora de "discrepar sobre la interpretación de las medidas u observaciones, la aceptación de hipótesis científicas por consenso,…".*

De idéntica forma se expresa en cuanto a la incapacidad predictiva de la astrobiología:

*"The predictive approach may not work if a system is too complex to allow predictions to be made, is too large, or operates over too long time scale to allow experiments to be performed or observed, or if random events that inherently cannot be predicted have a large effect on the outcome [...] On the biological side, again we cannot use a first principles approach to understand the path of evolution during the history life on Earth"*

Lo que le hace concluir que la naturaleza científica de la astrobiología está relacionada con las ciencias de tipo histórico-narrativo:

*"In dealing with the actual events that have already taken place in a system, we are looking at the past rather than making predictions about the future. We can look for evidence*

---

[13] Clark, K. Civilización. Madrid. Alianza Editorial. 2007 p.5
[14] Jakosky, B. Science, Society, and the search for life in the Univers. The University of Arizona Press.2007. pp. 65

*left behind by the past events and use it to infer what those events must have been. In deciphering the sequences of events from this evidence we are engaging in 'historical science'.*"[15]

La astrobiología sería pues una ciencia histórica que, con el objetivo de describir la historia de los entes vivos y sus precursores, utiliza los métodos propios de las ciencias *nomotéticas* como la física o la química que pueden conducir a predicciones o retrodicciones. Curiosamente, el objeto de estas últimas es el descubrimiento de leyes generales derivadas de la historia de casos particulares. Es decir, existe una suerte de relación especular entre las ciencias nomotéticas y las históricas que podría conducir a argumentaciones de tipo circular indeseables.

Otra forma de aproximación a la epistemología y métodología de la astrobiología es redactar un sumario de preguntas que nos aclaren el contenido de su disciplina:

- ¿Qué es la vida?
- ¿Qué procesos físicos, químicos y biológicos están relacionados con el origen de la vida?
- ¿Cómo y por qué emergió en el planeta Tierra?
- ¿Cuáles son las condiciones para que la vida emerja y persista en cualquier lugar?
- ¿Existe o existió realmente en algún lugar fuera de la Tierra?
- ¿Fue consecuencia necesaria de la evolución del Universo?

Todas estas cuestiones presentan un eje común que la filosofía denomina: "el problema de la vida", donde cuatro son los conceptos relacionados: la reducción, la emergencia, la complejidad y la direccionalidad.

De lo dicho hasta el momento se sigue que una de las cuestiones más importantes por dilucidar es el estatuto gnoseológico de la disciplina. Para intentar resolver este rompecabezas se requiere de un análisis conceptual, epistemológico y metodológico que nos oriente y aclare la forma en que la astrobiología afronta las soluciones a varios problemas: el origen, perseverancia y evolución de la vida en el Universo, el papel representado por la química, la física y la tecnología y de qué manera responde a un programa reduccionista y teleológico.

En lo que sigue, plantearemos el problema de la vida y analizaremos las influencias que las distintas posiciones filosóficas ejercen sobre la astrobiología. Posteriormente haremos una aproximación superficial sobre algunas cuestiones ontológicas y epistemológicas derivadas de la interdisciplinaridad: como el papel que desempeñan en el reduccionismo de corte fisicalista y la teleología.

---
[15] Ibd., p.73

### 3. La astrobiología y el problema de la vida:

La astrobiología representa un ejemplo paradigmático de cómo la ciencia actual está en continuo diálogo con la tradición científica clásica. El origen de la vida desde la materia inerte fue tratado por el atomismo[16] griego apoyado firmemente en presupuestos ontológicos que andando los años siguen siendo válidos para el fisicalismo y mecanicismo actual. Por otro lado, la ontología aristotélica sostiene, entre otras, las posiciones holistas y finalistas de algunos de los científicos y filósofos de la biología de nuestros días.

Aunque como nos dice el profesor González Recio:

*"La Vida no podía ser objeto de explicación puesto que cualquier explicación tenía en ella sus raíces; la Vida no fue inicialmente una región de la Naturaleza sino una manera de pensar la Naturaleza"*[17]

Nos resume las diferencias esenciales en los rasgos sobre la concepción del problema de la vida entre atomistas y peripatéticos[18]:

Atomistas:

1. Relevancia de la causalidad externa
2. Carácter accidental de la forma biológica
3. Ausencia de la idea de organismo como identidad irreductible
4. Explicación de la macroestructura a partir de la microestructura
5. Las plantas y los animales son efectos mecánicos surgidos en la historia accidental de la naturaleza, desprovistas de leyes morfogenéticas.

Peripatéticos:

1. Relevancia de la causalidad interna
2. Carácter esencial de la forma biológica
3. El organismo como unidad irreductible
4. explicación de la microestructura a partir de la macro estructura

---

[16] González Recio, José Luis. Aire, calor y sangre o la vida inventada desde el Mediterráneo. En: González Recio, J.L. (Ed.). Átomos, almas y estrellas. Plaza y Valdés. 2007.pp.148-200
[17] Ibd. p.147.
[18] Ibid. pp.175-177

5. Las plantas y los animales son productos teleológicos.

Estas continúan siendo las cuestiones filosóficas principales que dificultan el consenso de las distintas ciencias en cuanto a la naturaleza de los seres vivos.

El nacimiento de la astrobiología ha reactivado la preocupación científico-filosófica por encontrar una definición de vida gracias, sobre todo, a la necesidad práctica de diseñar dispositivos que puedan evidenciar la existencia de vida extraterrestre.

Una de las restricciones más importantes para afrontar la investigación sobre la vida en el Universo surge de la determinación de sus características, para lo cual la astrobiología se plantea la necesidad o futilidad de definir con precisión la naturaleza de la vida.

Las propiedades de los sistemas vivos se pueden resumir en ocho: la vida muestra orden y estructura, toma nutrientes y excreta deshechos, utiliza energía, crece y se desarrolla, lleva a cabo reacciones bioquímicas específicas, responde a su entorno, se reproduce y se adapta a las condiciones del ambiente[19]. Como veremos más adelante, todas estas son propiedades que pueden o no presentarse en algunos sistemas que calificamos como vivos, por lo que para cada caso existen excepciones tanto positivas – cumple la propiedad pero no es un sistema vivo – como negativas – no cumple la propiedad pero lo reconocemos como vivo-.

Cualquier teoría sobre la vida debe poder explicar los rasgos de la vida, los casos fronterizos y los denominados rompecabezas de la vida[20].

En lo que respecta a los rasgos distintivos de la vida, sus condiciones necesarias y suficientes son: el metabolismo que trata el procesamiento de la energía y los nutrientes – capaz de distinguir vivo, durmiente, muerto e imposibilidad de estar o haber estado vivo-, la homeostasis interna en un ambiente variable, la capacidad de portar información para posibilitar la reproducción – información, mutación, presión selectiva del entorno-.

Ejemplos de casos fronterizos son los autorreplicadores metabólicamente dependientes (virus, priones, parásitos, etc.), los superorganismos, la vida artificial débil (i.e el software 'Tierra').

Por último, los rompecabezas de la vida más importantes son: el origen, la emergencia, jerarquización, la continuidad o discontinuidad desde lo no vivo y la vida artificial fuerte. Estos son, sin duda, problemas abordados por la astrobiología. En concreto, el origen de la vida, su procedencia gradual de sistemas no vivos o por el contrario su emergencia como salto cualitativo son problemas biológicos que tienen sólidas propuesta como el "mundo del RNA"[21], la teoría de la endosimbiosis seriada y simbiogénesis de Lynn Margulis, etc.

---

[19] Jakosky, B. *Science, society, and the search for life in the Univers.* The University of Arizona Press. Tucson. 2006. pp. 30-45
[20] Idem. p.459

Dejando a un lado las corrientes vitalistas, la filosofía aborda el problema de la vida desde distintas posiciones ontológicas, metodológicas y epistemológicas: biosistemismo, escepticismo, funcionalismo, mecanicismo fisicalista y maquinismo, esencialismo, antiesencialismo, etc.

Para el esencialismo la vida no es un tipo de sustancia química fija, las características de la vida como el metabolismo, la reproducción, etc. trabajando al unísono explican las potencialidades causales evidenciadas mediante los experimentos y, por lo tanto, es en estas potencialidades donde reside la esencia de los tipos naturales.

Por el contrario, los antiesencialistas piensan que las propiedades de los sistemas vivos son azarosas, procedentes de una colección abstracta de todas las características posibles, pero a la vez necesarias y suficientes. Por eso desestiman los casos fronterizos. Algunos antiesencialistas argumentan que el abandono de las viejas taxonomías y por lo tanto el nacimiento de la concepción esencialista es consecuencia del debate abierto hace doscientos años sobre la posibilidad de crear vida de lo no vivo. Por lo tanto, la propuesta taxonomista es considerar que la vida tal como la conocemos es la que es y puede ser[22] y eso significa que tanto la vida en el laboratorio como la extraterrestre quedarían fuera de esta categoría.

Otras propuestas sobre la naturaleza de la vida se muestran escépticas y afirman que la biología puede continuar sin preocuparse por una definición de vida (Sober) y creen que ésta no puede categorizarse a través de condiciones necesarias y suficientes. La forma de interaccionar con el entorno (metabolismo), autoorganización, el almacenaje y la utilización de la información para la reproducción o la pertenencia a un grupo con la capacidad de evolución, representan meros aires de familia.

Algunos se aproximan al problema desde una orientación lingüística buscando una definición que especifique el significado de los términos. Concluyen que encontrar una definición es cuestión de convención lingüística porque los términos están categorizados por el hombre pero sólo la naturaleza determina su pertenencia. Por eso afirman que hasta que no encontremos una buena teoría de la naturaleza de los sistemas vivos seguirán existiendo ambigüedades procedentes de especificar un conjunto de propiedades que nos informan sobre el parecer y no sobre el ser. En palabras de Clelland y Chyba:

---

[21] Lehninger et al. *Principios de Bioquímica.* Ediciones Omega. 2007. pp. 74-78
[22] Idem. p. 462

*"... insights gained from phylosophical investigations into language and logic strongly suggest that the seemingly interminable nature of the controversy over life's definition is inescapable as long as we lack a general theory of the nature of living systems and their emergence from the physical world."*[23]

El ejemplo paradigmático que utilizan en sus argumentos es el caso del agua: hasta que la ciencia no encontró una teoría molecular de la materia que empíricamente comprobara la naturaleza del agua, la definición de agua representaba una convención lingüística que describía el conjunto de propiedades que la caracterizaban. Por lo tanto, para estos autores, todavía nos encontramos en el momento de generar conjeturas sobre la naturaleza de los sistemas vivos y someterlas a juicio empírico.

De lo expuesto hasta ahora, no cabe duda de que se pueden extraer consecuencias interesantes para el programa científico de la astrobiología, pero tal vez los sistemas filosóficos que se presentan más radicales, críticos e influyentes son respectivamente el biosistemismo y sus contrapartidas filosóficas, el fisicalismo y funcionalismo.

Para el organicismo o biosistemismo "'vida' es la extensión del predicado 'está vivo'[...] es una colección, por lo tanto, un objeto conceptual"[24]. De esto se deduce que "estar vivo" es una propiedad emergente de algunos sistemas complejos. Emergencia entendida desde el punto de vista ontológico como aquella propiedad del sistema de la que carecen todos sus componentes. Esta definición no tiene que ver con lo que conocemos o ignoramos de las cosas, es decir, con su episteme, sino con ellas mismas. Por lo tanto, reconoce la vida como un nivel emergente surgido de lo químico[25]. En este sentido la célula es el biosistema elemental, aunque no todas las células son biosistemas elementales. Es decir, la célula es el sistema mínimo donde emerge el $βιος$ en un tiempo determinado y esta emergencia es un cambio cualitativo respecto a cualquier sistema prebiótico. Esto es lo que caracteriza la ontología de la propuesta. Por lo tanto descartan la vida artificial en sentido fuerte (VAF) ya que las propiedades emergentes del $βιος$ son inseparables de las cosas en sí por lo que la VAF crea un claro problema ontológico. Además, los sistemas artificiales sólo estarían vivos por definición lo que supone un problema epistemológico.

En lo que respecta a los casos fronterizos: los virus, los priones, el RNA, las mitocondrias, etc. pertenecen a sistemas moleculares que no son biosistemas. Esto es así porque propiedades como la replicación, el metabolismo, etc. no son propiedades necesarias y suficientes para calificar al sistema molecular como biosistema.

---

En la propuesta materialista biosistémica emergentista, los organismos en su ambiente con sus subsistemas (moléculas, células y órganos) y supra sistemas (población, ecosistema, etc.) son las unidades de la biología[26].

Se pueden extraer varias consecuencias importantes para la astrobiología: los biosistemas así definidos sólo se dan en la Tierra por lo que la exobiología no tiene objeto científico y es meramente especulativa[27], el origen (en laboratorio, o azaroso por autoorganización prebiótica) y la historia ( recordemos que la capacidad de evolucionar no es propiedad necesaria ni suficiente) de los biosistemas no dice nada importante sobre la propiedad de "estar vivo" y, por lo tanto, no podemos concluir nada de las investigaciones sobre vida artificial en sentido fuerte. En estos planteamientos, los biosistemistas coinciden con la postura epistémica de los taxonomistas para los que la vida como conjunto de las cosas vivas es la vida tal como la conocemos en la Tierra.

No obstante, se puede defender la astrobiología junto con las ciencias de la astrofísica y la astroquímica dentro de un programa de investigación reduccionista débil: la reducción como método para analizar los distintos componentes y propiedades químico-físicas del biosistema, que caracterizan los sistemas moleculares de los cuales emergerán los biosistemas. También puede ser de utilidad desde el biosistemismo emergentistas la cualidad de interdisciplinaridad que presenta la Astrobiología, ya que la unificación o integración es otro método gnoseológico alternativo a la reducción. Pero se debe tener en cuenta que no todas las teorías son unificables[28] - pues deben compartir conceptos y referentes -funciones, variables, hipótesis, métodos, preguntas,…- y fórmulas puente que aglutinen y cohesionen las distintas ciencias.

La escuela filosófica enfrentada al biosistemismo emergentista es la reducción fuerte propuesta por el fisicalismo - también en su versión maquinista- y el programa funcionalista.

Para los fisicalistas, los organismos son sistemas fisicoquímicos complejos pero regidos por las leyes físicas y químicas. Por lo tanto sus propiedades no requieren de argumentos fuera de la física y la química. El maquinismo afirma la naturaleza mecánica de los organismos.

La propuesta reduccionista afirma que la bioquímica es "universal" y que son suficientes las leyes de la física para dar cuenta de todos los fenómenos moleculares. De ello se deduce que la biología molecular y la bioquímica son dos nombres para una misma ciencia: la física de las moléculas. Es decir, sería suficiente determinar los límites termodinámicos, energéticos, materiales y geográficos para establecer las posibilidades de la vida.

---

[26] Idem. p.173-175
[27] Idem. p.169
[28] Idem. p.142

Como ejemplo reciente de la posición reduccionista fuerte son clarificadoras las palabras del físico Esteen Rasmusen ante la primera "célula sintética" conseguida en el laboratorio de Craig Venter con el que discrepa en cuanto a lo que se entiende por vida:

> *"Bottom-up researchers, such as myself, aim to assemble life — including the hardware and the program — as simply as possible, even if the result is different from what we think of as life. All of these deeply entrenched metaphysical views are cast into doubt by the demonstration that life can be created from non-living parts, albeit those harvested from a cell"*[29]

La posición funcionalista comparte el objetivo de conseguir 'vida artificial' pero su programa es *top-down*: parte de los biosistemas como la célula para describir las propiedades de lo viviente mediante, por ejemplo, el cambio del programa genético que opera el *hardware* de la célula. Es decir, pretende sintetizar las partes del sistema viviente y luego organizarlas para que desarrollen el comportamiento dinámico de la vida. De esta forma su hipótesis es que todas las formas de vida son similares a la de la Tierra y descarta otros metabolismos, sistemas de mantener y reproducir la información, etc. Las 'células artificiales' mediante el ensamblaje, la codificación de la información sobre el sistema –y sus errores- y el metabolismo nos recrean los sistemas autocontenidos que intercambian energía con el exterior y evolucionan.

El propio Craig Venter nos reduce todo a lo que denomina genoma mínimo:

> *"minimal genome sufficient to support life"*[30]

Los planteamientos funcionalistas explican algunas propiedades pero fallan en otras. Por ejemplo, los basados en el metabolismo son capaces de explicar problemas como la vida en estado 'durmiente' pero descartan como vivientes otros sistemas que presentan propiedades metabólicas (la llama, las células convectivas, la transformación del hierro en óxido de hierro, etc.). Lo mismo sucede con los funcionalistas que definen la vida en términos de la autoorganización y auntomantenimiento –autopoyesis-, aunque sabemos que se pueden encontrar sistemas no vivos con estas cualidades (cristales, etc.).

---

[29] Rasmusen, S. *'Bottom-Up' will be more telling.* Nature. 25-05-2010. vol. 456. p.422.
[30] Apud. Bedau, M.A. *What is life?.* En: "Sarkar, S. *A Companion to the Phylosophy of Biology.* Blacwell Publisher. 2007. p.463

## 4. Astrobiología: entre transdisciplinaridad e interdisciplinaridad

Bajo este debate fisicalismo-funcionalismo-biosistemismo se encuentra el problema de la reducción que en palabras de Rosenberg:

*"...es una tesis metafísica, una postura acerca de la explicaciones, y un programa de investigación"*[31]

Aunque las tesis fisicalistas y funcionalistas coinciden con las biosistemistas en que hay un sustrato físico-químico de los sistemas vivos, se contraponen en la interpretación sobre las potencialidades causales de de las propiedades biológicas. Para los primeros las potencialidades causales biológicas están determinadas por las partículas, atomos, moléculas y sus interacciones. Los antireduccionistas sostienen que existe un principio de realidad autónoma de lo biológico irreductible a la biología molecular que implica una primacía explicativa para la cual "al menos en ocasiones los procesos en el nivel funcional proveen la mejor explicación para los procesos en el nivel molecular"[32]

Si la biología no puede ser reducida fuertemente a la física, entonces los principios físicos universales no encuentran continuidad en una hipotética universalidad de las leyes de la biología. El asunto es que parece improbable esa reducción fuerte de la biología ya que ni siquiera la propia física reduce fuertemente las leyes físicas del macrocosmos al microcosmos. Es decir, las leyes puente tendidas desde la física del microcosmos –la mecánica cuántica- son necesarias pero no suficientes para construir las leyes del meso y macrocosmos. Se deben añadir hipótesis fuera de la física cuántica para poder explicar la biología – por ejemplo el principio de complementariedad, etc.-.

De esta manera, si la reducción es de tipo débil, la física no reduce a la química y ésta no puede reducir a su vez a la biología, por lo que no es determinante sobre los papeles y funciones biológicas de las moléculas en los biosistemas. De esto se concluye que las células, los organismos, etc. son sistemas con una estructura y organización que transcienden las propiedades de los demás sistemas químicos.

El fisicalismo fuerte y el funcionalismo se contraponen al biosistemismo de forma que sus consecuencias ontológicas, epistémicas y metodológicas se enfrentan favorecidas por la transdisciplinaridad de la astrobiología. Se entiende transdisciplinaridad como sinónimo de multidisciplinaridad, es decir, muchas ciencias trabajando unas de espaldas a otras. Ciencias

---

[31] Rosenberg, A. *Reductionism in a Historical Science*. Philosophy of Science 68 (2):135-163.2001 p.135
[32] Rosenberg, A. *Reductionism redux: computing the embryo*. Biology and Philosophy. 1997 p.446

que utilizan sus propios métodos, definen sus propios conceptos, establecen sus propias leyes, fundamentos y problemas. Sin diccionario que oriente, unifique, traduzca y defina. Sin prácticas empíricas que puedan fundir metodologías sobre las que poder establecer leyes aglutinantes.

Pero con más precisión nos debemos preguntar por el papel que juega la astrobiología en el debate entre fisicalismo, funcionalismo y biosistemismo. La transdisciplinaridad puede ser parte del problema a resolver por los astrobiólogos.

La astrobiología se aproxima al problema de la vida desde dos orientaciones: la orientación abajo-arriba, es decir, una orientación causal ascendente como la propuesta por el fisicalismo y una orientación de arriba-abajo, es decir, una orientación causal descendente seguida por biosistemismo.

Las disciplinas que tratan el modelo causal ascendente son la física y la química en las subdisciplinas de la astrofísica, astroquímica y la cosmología. La historia de la materia propuesta por la cosmología –basada en la mecánica cuántica y en la relatividad general- propone un proyecto reduccionista fuerte de tipo teleológico apoyado en la cosmología estándar. De los tres primeros minutos desde el comienzo del Universo hasta el presente la materia se fue transformando según se desacoplaban las cuatro fuerzas fundamentales del Universo: gravitación, nuclear fuerte, nuclear débil y electromagnética. Conforme el Universo se enfriaba, se produjo la aparición de las partículas elementales y la energía a las que siguieron los átomos de H, He y Li que se aglutinaron para formar las estrellas gigantes de primera generación, cuya evolución y muerte en forma de supernova enriqueció el medio interestelar con el resto de elementos pesados que forman la tabla periódica. Este medio interestelar enriquecido y enfriado dio lugar a moléculas complejas entre las que se encuentran moléculas orgánicas como los hidrocarburos policíclicos aromáticos detectados por los telescopios de infrarrojos. El final de esta narración de la 'historia de la materia', necesariamente direccional, continua, tendente a la complejidad, tiene como antepenúltima parada la biología molecular que con suavidad se moverían desde el mundo del RNA a las pleneuromas con penúltima parada en los sistemas vivos y con destino final en la autoconciencia. En la propuesta de Stuart Kauffman:

> *"El resultado es un apoyo sin ambages a la teoría de que el proceso evolutivo es la consecución de un proceso de autocatálisis de complejidad creciente matizado- ¿por qué no?- por la evolución"* [33]

Hay evidentemente muchos problemas en esta narración entre ellos se encuentra el de conseguir "vida natural" además de la autodefinida "vida artificial".

---
[33] Castrodeza, C. *Los Límites de la Historia Natural*. Akal. 2003 pp. 16-17

Por lo demás, el problema de la complejidad tiene muchos tentáculos no resueltos desde el fisicalismo: por ejemplo la diferencia que muestra el papel desempeñado por la termodinámica de los procesos irreversibles, que son los más comunes en biología, comparada con la de los fenómenos reversibles que ha sido tratada por la física con mayor detenimiento. Es interesante pensar en las consecuencias que esta cuestión puede tener en el debate entre neutralistas y direccionalistas. Un debate que sin duda está ligado al del Universo como sistema cerrado que defienden algunas propuestas cosmológicas frente al Universo semi-abierto de los sistemas biológicos. Da que pensar al respecto que la mayoría de los astrobiólogos, o con más propiedad cosmólogos, como Juan Pérez Mercader se consideren direccionalistas.

Por otro lado, el modelo causal descendente también denominado filogenia molecular pretende extrapolar los sistemas vivos hacia épocas próximas al origen de la vida en la Tierra. El árbol filogenético hunde sus raíces en el remoto pasado donde señala la presencia de un ancestro común hace 3,8 millones de años. Sin duda ésta es una de las concepciones heredadas de la *Naturphilisophie* alemana[34]. Una de las consecuencias más interesantes te este viaje en el tiempo es que la mayoría de los organismos próximos a la raíz del árbol no utilizan la luz como fuente de energía. Esta es la razón por la que se trabaja con la hipótesis de que el antepasado común muestra grandes similitudes con los organismos quimiosintéticos que pueblan las fumarolas hidrotermales escondidas bajo el mar. Es decir, el árbol de la vida basado en los datos del ribonosoma RNA tiene su raíz constituida por organismos termófilicos e hipertermofílicos. Detrás de esta teoría se encuentran las leyes de la evolución con el correspondiente problema de la reducción entre la biología molecular, la biología del desarrollo y la biología de la evolución.

A primera vista resalta que las dos aproximaciones se muestran difícilmente reconciliables. Por ejemplo, está claro que la aproximación hacia abajo no nos puede informar de tipos de vida distintos a los terrestres. Por lo que su ontología, epistemología y metodología se adscribirá con más facilidad al biosistemismo. De igual manera podemos afirmar que la dirección ascendente del fisicalismo y del funcionalismo puede definir la vida a su gusto y por lo tanto puede ser de una naturaleza distinta a la vida terrestre. Como hemos comentado más arriba, la computación aplicada al diseño de vida artificial fuerte es una de las apuestas punteras del funcionalismo.

A modo de conclusión, estimamos que la clave del problema de la astrobiología como disciplina científica reside en la necesidad de encontrar problemas comunes y teorías compatibles con todas las ramas de la ciencias implicadas. Conceptos como emergencia, complejidad y evolución no son equivalentes en las distintas disciplinas científicas. Por lo que se generan debates estériles y mucha confusión. Por lo tanto, es imprescindible conseguir una cierta interdisciplinaridad en astrobiología y consideramos importante el papel que puede jugar la filosofía para desenredar la madeja tejida, en parte, por la transdisciplinaridad.

---

[34] Ibd. p.20